\documentclass[superscriptaddress,floatfix,showpacs,pra,aps,twocolumn]{revtex4-1}

\usepackage{bm}
\usepackage{amsmath}
\usepackage{amssymb}
\usepackage{latexsym}
\usepackage{amsfonts}
\usepackage{graphicx}
\usepackage{blindtext}
\usepackage{dsfont}
\usepackage[english]{babel}
\usepackage[usenames,dvipsnames,svgnames,table]{xcolor}

\newcommand{\ket}[1]{\left|#1\right>}

\newcommand{\braket}[2]{\left<#1|#2\right>}
\newcommand{\bin}[2]{\begin{pmatrix}#1\\#2\end{pmatrix}}
\newcommand{\sbin}[2]{\left(\begin{smallmatrix} #1 \\ #2 \end{smallmatrix}\right)}
\newcommand{\ID}{\ensuremath{\mathds{1}}}
\renewcommand{\H}{\ensuremath{\mathcal H}}
\newcommand{\Hi}{\ensuremath{{\mathcal H}_{i}}}
\newcommand{\Hf}{\ensuremath{{\mathcal H}_{f}}}

\renewcommand{\d}{\ensuremath{\mathrm{d}}}

\newcommand{\ord}[1]{{\cal O}(#1)}
\newcommand{\abs}[1]{{\left| #1 \right|}}

\begin{document}

\title{Probing nonlinear adiabatic paths with a universal integrator}

\author{Michael Hofmann}
\email{michael.hofmann@wias-berlin.de}
\affiliation{Weierstra{\ss}-Institut f\"ur Angewandte Analysis und Stochastik,
Mohrenstra{\ss}e 39,
D-10117 Berlin, Germany}
\affiliation{Institut f\"ur Theoretische Physik, 
Technische Universit\"at Berlin,
Hardenbergstra{\ss}e 36, D-10623 Berlin, Germany}
\author{Gernot Schaller}
\affiliation{Institut f\"ur Theoretische Physik, 
Technische Universit\"at Berlin,
Hardenbergstra{\ss}e 36, D-10623 Berlin, Germany}

\begin{abstract}
We apply a flexible numerical integrator to the simulation of
adiabatic quantum computation with nonlinear paths.
We find that a nonlinear path may significantly improve the 
performance of adiabatic algorithms versus the conventional
straight-line interpolations.
The employed integrator is suitable for solving the time-dependent 
Schr\"odinger equation for any qubit Hamiltonian.
Its flexible storage format significantly reduces cost for storage and 
matrix-vector multiplication in comparison to common sparse matrix schemes.
\end{abstract}


\maketitle

%%%%%%%%%%%%%%%%%%%%%%%%%%%%%%%%%%%%%%%%%%%%%%%%%%%%%%%%%%%%%%%%%%%%%%%%

Simulating quantum systems requires enormous computational resources: 
Even for a few hundred particles there would be more variables to be stored than atoms 
exist in the universe~\cite{stolze2003}. 
To turn this problem into an advantage, quantum computers may be efficiently used for 
such simulations, since they are quantum systems themselves~\cite{feynman1982a}. 
Moreover, quantum algorithms can solve distinct problems like number factoring with 
exponential speedup compared to classical computers~\cite{shor1997a}. 
 
In the conventional picture, quantum algorithms are implemented as a sequence of
unitary operations~\cite{nielsen2000}, which implies fast switching of the generating Hamiltonian.
In contrast, within the paradigm of adiabatic quantum computation~\cite{farhi2001a}, the Hamiltonian
is modified slowly from a simple initial Hamiltonian with an easy-to-prepare ground state to a final 
Hamiltonian which encodes in its ground state the solution to some difficult problem. 
Most importantly, for a large class of problems, implementation of the final Hamiltonian is possible
without knowing the solution of the problem explicitly.
The adiabatic theorem implies -- provided the evolution is slow enough -- that the system will end
up near the ground state of the final Hamiltonian, such that the solution to the problem can be
obtained by measuring the system.
The evolution time is related to the spectral properties of the time-dependent Hamiltonian and
thus corresponds to the algorithmic complexity of an adiabatic quantum algorithm (AQA).
The conventional circuit picture and the adiabatic approach are known to be polynomially 
equivalent~\cite{aharonov2007a,mizel2007a}, but exact results for adiabatic algorithms are scarce~\cite{roland2002a}.
It is therefore quite interesting that first numerical simulations of the Schr\"odinger equation revealed
a seemingly polynomial complexity of the adiabatic algorithm for an NP-complete problem~\cite{farhi2001a}.
Since then, it has been a strongly debated question whether this scaling would persist for larger problem 
sizes~\cite{boulatov2003a,znidaric2006a,ioannou2008a,young2008a,amin2009c,schaller2010a}.
Recent findings suggest that the scaling complexity of the conventional straight-line
adiabatic interpolation is typically exponential~\cite{altshuler2010a,young2010a}.
It may however be conjectured that with modifications of the adiabatic algorithm, its scaling behavior
can be considerably improved~\cite{farhi2008a}, such that the scaling behavior of adapted algorithms
is still an open question.

Unfortunately, this question can currently not be settled from the experimental side:
Though enormous progress has been made in the last decade, not more than  
a few quantum bits (qubits) have been entangled so far~\cite{monz2011a}, which 
currently restricts the execution of quantum algorithms to proof-of-principle demonstrations.
As experiments are still neither flexible nor scalable enough to investigate new theoretical models, 
the demand for classical computer simulations of quantum algorithms is growing. 
Such simulations are computationally expensive and usually must be coded separately for each
problem considered.
Here, we use an efficient numerical integrator to solve the time-dependent Schr\"odinger equation 
for the high-dimensional but sparse Hamiltonians typical for qubit systems. 
An adopted storage format will reduce the memory required for storing the Hamiltonian in comparison to 
common sparse matrix schemes while keeping their advantage of fast matrix-vector multiplication.
In particular its ability to follow flexible adiabatic paths renders our storage scheme suitable
for such simulations.

The paper is structured as follows:
In Sec.~\ref{sec:theory}, we expose the prerequisites discussing the data storage scheme, adiabatic computation, and the particular
NP-complete problem considered.
Afterwards, we numerically compare the performance of different adiabatic quantum algorithms (AQAs) for
straight-line interpolation in Sec.~\ref{sec:straight}.
Then, we turn to the investigation of non-linear paths in Sec.~\ref{sec:alternative} and close with conclusions.

%%%%%%%%%%%%%%%%%%%%%%%%%%%%%%%%%%%%%%%%%%%%%%%%%%%%%%%%%%%%%%%%%%%%%%%%

\section{Theory}
\label{sec:theory}
\subsection{Sparse Quantum Hamiltonian (SQH)}
%%%%%%%%%%%%%%%%%%%%%%%%%%%%%%%%%%%%%%%%%%%%%%%%%%%%%%%%%%%%%%%%%%%%%%%%

A single-qubit state is a superposition of two fundamental states denoted by $\ket0$ and $\ket1$, which form the computational basis. As a convention, those states are the eigenstates of the Pauli matrix $\sigma^z$: $\ket0 := \sbin{1}{0},\; \ket1 := \sbin{0}{1}$. Similarly, the basis states for an $n$-qubit system can be constructed by the tensor product
\begin{equation}
  \label{eq:ztensor}
  \ket z = \bigotimes_{i=1}^n \ket{z_i},\;z_i\in\{0,1\}.
\end{equation}
Here $z$ is the decimal representation $z=\sum_{i=1}^n z_i\,2^{i-1}$ of the bitstring $z_n\,z_{n-1}\ldots z_2\,z_1$. An arbitrary $n$-qubit state is then given by the superposition
\begin{equation}\label{eq:expansion}
  \ket\psi = \sum_{z=0}^{2^n-1} \alpha_z\ket z,\; \alpha_z \in \mathds C
\end{equation}
with normalization condition $\sum_z \left| \alpha_z\right|^2 = 1$. Obviously, the dimension of the Hilbert space, $N=2^n$, is growing exponentially with the number of qubits which makes simulations of quantum systems hard.

A Hamiltonian acting on $n$ qubits can be described by the Pauli matrices $\sigma^x,\sigma^y,\sigma^z$: Together with the identity $\sigma^0:=\ID$ they span the space of all $2\times2$-matrices. Using the $n$-fold Kronecker product of those matrices yields $N^2=2^{2n}$ generalized Pauli matrices (GPMs) $\mathcal S_i= \bigotimes_{q=1}^n \sigma^{\alpha_{i,q}}$ as a basis for all $N\times N$-matrices. 
Trace-orthogonality ensures that any Hamiltonian can be decomposed into GPMs, 
\begin{equation}
  \label{eq:H2GPMs}
  \mathcal H = \sum_{i=0}^{N^2-1} m_i \,\mathcal S_i
\quad\text{with}\quad m_i = \frac{1}{N}\mathrm{Tr}\{\mathcal H\,\mathcal S_i\} \in \mathds R. 
\end{equation}
Let $\sigma_i^\alpha$ be the short notation of the tensor product
\begin{equation}
\underbrace{\ID\otimes\ldots\otimes\ID}_{i-1}\otimes\;\sigma^\alpha\otimes\underbrace{\ID\otimes\ldots\otimes\ID}_{n-i}, \qquad \alpha \in \{x,y,z\},
\end{equation}
acting only on the $i$-th qubit. A GPM of order $j$ (acting non-trivially on $j$ qubits) is then written as  $\sigma_{i_1}^{\alpha_1}\sigma_{i_2}^{\alpha_2}\ldots\sigma_{i_j}^{\alpha_j}$. 
%Furthermore, 
By counting only non-vanishing terms $m_i\ne 0$, Eq.~(\ref{eq:H2GPMs}) has the more convenient form
\begin{equation}
  \label{eq:SQH-Hamiltonian}
  \mathcal H = m^{(0)}\ID + \sum_{j=1}^p\sum_{l=1}^{k_j} m_l^{(j)} \prod_{q=1}^j \sigma_{i_{l,q}^{(j)}}^{\alpha_{l,q}^{(j)}}\,,
\end{equation}
%GS
where $p$ denotes the order of the Hamiltonian, $k_j$ the number of $j$-local terms,  $m_l^{(j)}$ the corresponding real prefactor, and  $m^{(0)}$ the energy shift.

We now introduce a \emph{Sparse Quantum Hamiltonian (SQH) format}: For a complete description of the Hamiltonian's structure, we do not need to store the full GPMs but only the parameters used in Eq.~(\ref{eq:SQH-Hamiltonian}), including the positions $i$ and types $\alpha$ of the (single) Pauli matrices.
Consequently, storage of a Hamiltonian is efficient if its order is independent of the number of qubits, 
$p\ll n$, and becomes even more favorable when the number of terms for every order $j$ is small, e.g.,  
$k_j\sim\ord n$. 
An example of such a system is the quantum Ising model in a transverse field,
\begin{equation}
  \label{eq:ising}%GS
 \H = -\Omega(1-s)\sum_i^n \sigma_i^x - \Omega s\sum_i^n \sigma_i^z\sigma_{i+1}^z\,,
\end{equation}
where $\Omega$ represents an energy scale and $s$ a control parameter and where
periodic boundary conditions are assumed $\sigma^z_{n+1} = \sigma^z_1$.
This Hamiltonian would only require $\ord n$ elements to store in SQH format. 

A matrix-vector product $\H\ket\psi$ is according to Eq.~(\ref{eq:SQH-Hamiltonian}) reduced to a  sum of GPMs acting on a basis state, $\sigma_{i_1}^{\alpha_1}\ldots\sigma_{i_j}^{\alpha_j}\ket z$. 
Due to the tensor structure %$\ket z=\ket{z_1}\otimes\ldots\otimes\ket{z_n}$
of $z$ (c.f.\ Eq.~(\ref{eq:ztensor})), such a multi-qubit operation is broken down to a multiplication of successive single-qubit operations $\sigma_i^\alpha\ket{z_i}$. Assuming the number of terms in \H\ is $\ord n$, the effort
%GS
for computing $\H\ket\psi$ scales as $N\times\ord n$ instead of $N^2$ when the Hamiltonian was stored conventionally.

The universal applicability of the SQH representation allows to write program code (e.g., an integrator) 
independent of the used Hamiltonian as long as it has the SQH structure given in Eq.~\eqref{eq:SQH-Hamiltonian}.
Also time-dependent Hamiltonians can be implemented by time-dependent coefficients $m_l^{(j)}(t)$.

\subsection{Adiabatic Quantum Computation}
\label{sec:applications}

Quantum computation by adiabatic evolution has been 
suggested as a promising approach for solving NP-complete problems~\cite{farhi2001a}.
The idea is simple: A final Hamiltonian \Hf\ is constructed which encodes the solution for a
computational problem in its ground state, e.g., as an energy penalty function. 
We stress here that for a number of hard problems this can be done without knowing the solution.
Starting from an easy to construct ground state of an initial Hamiltonian \Hi, the system is transformed to \Hf\ after the runtime $T$. A common approach is to use a linear interpolation:
\begin{equation}
  \label{eq:aqc}
  \H(s) = (1-s)\Hi + s\Hf
\end{equation}
with constant velocity $s(t)=\frac{t}{T}$ and $0\leq t \leq T$.
If this transformation is slow enough, the adiabatic theorem guarantees that the system 
remains always near the instantaneous ground state~\cite{sarandy2004a} such 
that a measurement after time $T$ yields the solution. 
This works in principle with every energy eigenstate of the system, but by using the ground state one hopes that the evolution is 
dissipation-free (which becomes relevant when the system is coupled to a low-temperature reservoir~\cite{childs2001a}). 
The runtime $T$ thus is a measure for the algorithmic complexity.
It can be optimized by adapting the speed of the interpolation $\dot{s}(t)$ to the energy gap above the ground state~\cite{roland2002a,jansen2007a,schaller2006b}.
In this case however, the time-dependent Hamiltonian is still a convex combination of initial and final Hamiltonian (hence the terminology straight-line interpolations), 
and we will not consider such extensions here.

To study the efficiency of such an algorithm, i.e., the runtime scaling with increasing system size $n$, 
we need to determine an adiabatic runtime $T_s(n)$  by simulating the evolution of a system 
prepared in the initial ground state. It is governed by the Schr\"odinger equation
\begin{equation}
  \label{eq:schroedinger}
  i\hbar\frac{\d}{\d t}\ket{\psi(t)} = \H{(t)} \ket{\psi(t)},
\end{equation}
which for the expansion in Eq.~(\ref{eq:expansion}) becomes a whole set of $N$ 
coupled ordinary differential equations with initial condition $\ket{\psi(0)}=\ket{\psi_0}$. 
A fast  numerical integration  is achieved by using the SQH format for $\H(t)$ and a 
fourth-order predictor-corrector scheme~\cite{gershenfeld2000}, 
which requires only a single evaluation of $\H\ket{\psi}$ per integration step.

\subsection{3-Bit Exact Cover}
\label{sec:ec3}

A common problem for probing adiabatic quantum algorithms (AQAs) is 3-bit exact cover (EC3), 
which is NP-complete. 
In a nutshell, solutions to problems in the class NP can be verified (with a classical computer) in a time that is polynomial in the length of their input.
The completeness property in addition implies that every other problem in NP can be mapped to EC3 with polynomial
overhead only.

\begin{figure}[h]
  \centering
  \includegraphics[width=7cm]{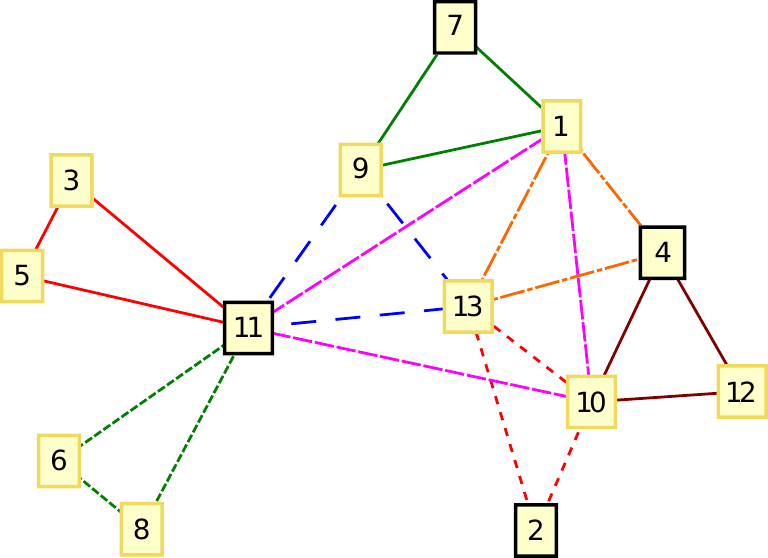}
  \caption{(Color Online) Instance of EC3, where the labeled vertices represent the 13 bits $z_i$,
    edges of same color and line style represent a clause $c_i$.  Bold-bordered vertices indicate
    the upbits in the solution $z=1098=(0010001001010)$.}
  \label{fig:ec3}
\end{figure}

On an $n$-bitstring $z_n\,z_{n-1}\ldots z_2\,z_1$ we define an instance of EC3 as a set of $m$ different clauses $c_i$ each involving  three different bits
\begin{equation}
  \label{eq:1clause}
  c_i = (c_i^{(1)},c_i^{(2)},c_i^{(3)}),\qquad c_i^{(k)} \in \{1\ldots n\}
\end{equation}
with $1\le i\le m$.
A clause is satisfied,
if and only if one of the involved bits $z_i$ equals 1, i.e., when
\begin{equation}
  \label{eq:sat}
  z_{c_i^{(1)}} + z_{c_i^{(2)}} + z_{c_i^{(3)}} = 1\,, 
\end{equation}
where '$+$' denotes the ordinary integer sum.
A solution to an instance is a bitstring satisfying all clauses in the set, which is easy to check. In contrast, finding such a bitstring is a combinatorial search problem for which no efficient classical algorithm is known. Figure \ref{fig:ec3} visualizes an EC3 instance for 13 bits with a unique solution.

For an AQA the problem is encoded as a cost function, where each unsatisfied clause adds
an energy penalty to the Hamiltonian~\cite{banuls2006a},
\begin{equation}
  \label{eq:hci}
  h_i = \left[ \ID - \frac{1}{2}\sum_k^3 \left(\ID-\sigma^z_{c_i^{(k)}}\right) \right]^2.
\end{equation}
The final Hamiltonian \Hf\ is simply constructed as a sum over all clauses and can be simplified to \cite{schuetzhold2006b}
\begin{equation}
  \label{eq:final_ham}
  \Hf = \sum_{i=1}^m \Omega h_i = \Omega m\ID - \Omega \sum_i^n \frac{n_i}{2}\sigma_i^z + \Omega \sum_{i<j}^n \frac{n_{i,j}}{2}\sigma_i^z\sigma_j^z.
\end{equation}
Here, $\Omega>0$ just denotes an energy scale, the coefficient $n_i$ denotes the number of clauses involving the i-th qubit,
and $n_{i,j}$ is the number of clauses, which contain the $i$-th and $j$-th qubit. 
For example, in Fig.~\ref{fig:ec3} we have $n_{11}=4$ and $n_{11,13}=1$. The Hamiltonian \Hf\ corresponds to a frustrated antiferromagnet in a non-uniform magnetic field (with an energy shift $m$) \cite{schuetzhold2006b}.
%GS
The coupling strength between the spins, however, is not defined by the experimental geometry 
(e.g., only between nearest neighbors) but by the edges of the clauses $n_{i,j}$, which
may define a highly disordered network.

\subsection{Hard Instances}
\label{sec:hard-instances}

To provide statistical evidence for our simulations, we generate for each system size 100 hard
instances.
These were characterized by a unique solution, a number of clauses 
close to the classical EC3 phase transition $m\approx\frac{2}{3}n$ from satisfiable to unsatisfiable 
problems~\cite{kalapala0508037}, and the constraint that $n_{ij} \in\{0,1\}$.
The last constraint implies that clauses should only share vertices and not edges.
It is motivated by the fact that in case an edge is shared by two clauses, 
one may easily conclude that in the solution, the opposite vertices
must have the same value.
Effectively, clauses that share edges would thus reduce the size of the problem.
Altogether, these constraints lead to $\ord n$ terms in Eq.~\eqref{eq:final_ham}, and \Hf\ is efficiently stored in SQH format.

\begin{figure}[htb]
  \centering 
  \includegraphics[width=0.9\linewidth]{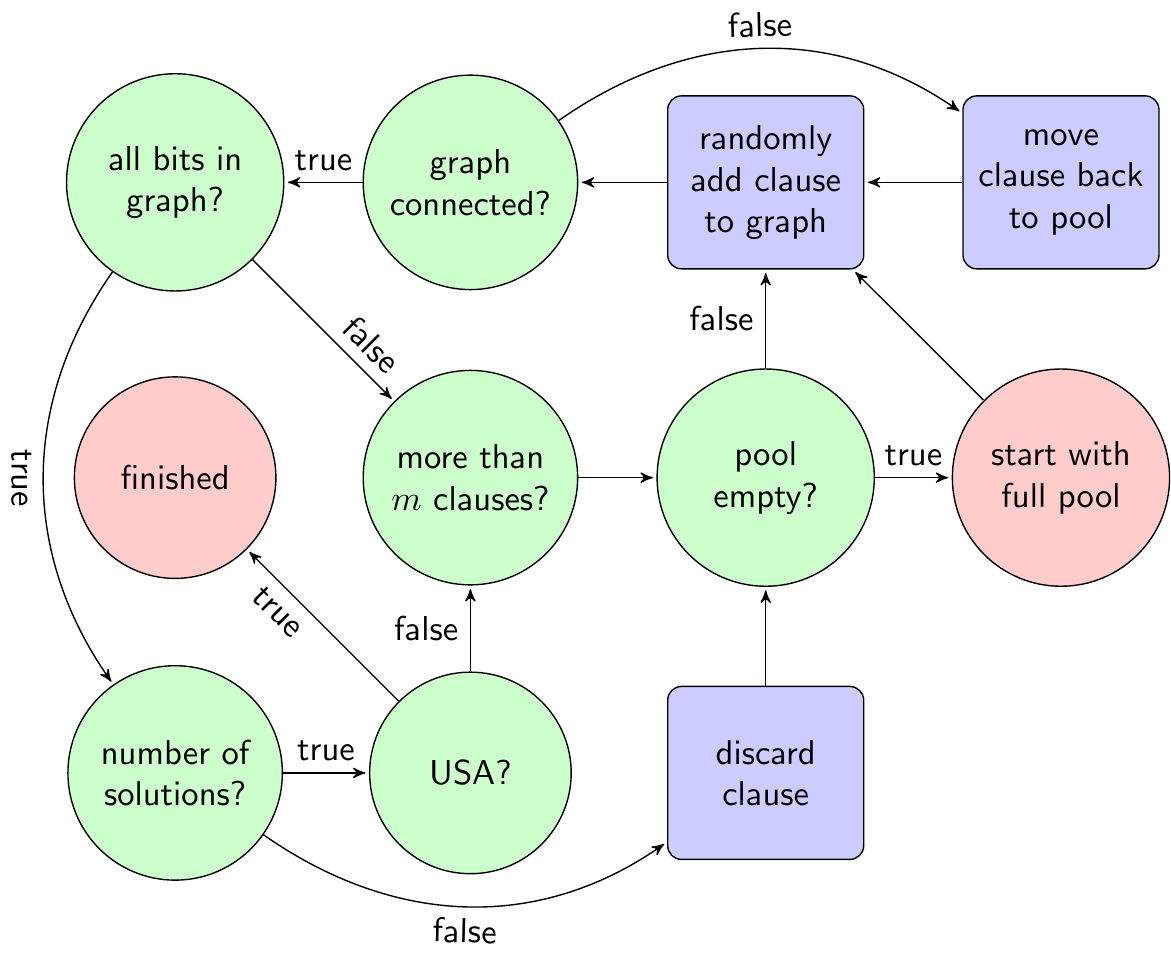}
  \caption{(Color Online) Scheme of the used algorithm for randomly generating hard instances of EC3. Rectangles
    and circles indicate instructions and if-clauses, respectively. The algorithm starts and
    stops at diamonds.}
  \label{fig:hardEC3}
\end{figure}

The recipe for generating a random (hard) instance is shown in figure \ref{fig:hardEC3}. 
We start with a full pool of all possible clauses. 
Randomly choosing one of the clauses defines an initial graph -- a
triangle. This graph is now iteratively increased by randomly choosing among the
remaining clauses whilst obeying simple rules: 
First, only clauses which intersect with the existing graph are added to ensure connectivity. 
When all bits are connected to the graph, the number of solutions is checked each time after a
clause is drawn. A clause is discarded, if it reduces the number of solutions to zero. The algorithm
finishes if only one solution is left. If no such unique satisfying assignment (USA) has been found
but all clauses are dropped from the pool \emph{or} the number of allowed clauses $m$ is reached,
the algorithm restarts itself.

%%%%%%%%%%%%%%%%%%%%%%%%%%%%%%%%%%%%%%%%%%%%%%%%%%%%%
\section{Straight Line Interpolation}
%%%%%%%%%%%%%%%%%%%%%%%%%%%%%%%%%%%%%%%%%%%%%%%%%%%%%

\label{sec:straight}

In this section, we simulate and compare the results of three  AQAs defined by  different initial Hamiltonians \Hi. 
The interpolation path between initial and final Hamiltonian is a straight line given by Eq.~\eqref{eq:aqc}, traversed at constant speed $s(t)=t/T$.
%GS
As a benchmark, we compare with the analytically solvable Ising model in Eq.~\eqref{eq:ising}, which
exhibits an inverse $1/n$-scaling of the minimum energy gap between ground and first coupled excited state, leading to
a quadratic scaling of the adiabatic runtime~\cite{dziarmaga2005a}.
We begin by summarizing the explored algorithms.

\subsection{Algorithms}

\subsubsection{X-Algorithm}

The original approach \cite{farhi2001a} used a single-qubit structure
\begin{equation}
  \label{eq:Hix}
  \Hi = \H^x = \Omega \sum_i^n \frac{n_i}{2} (\ID-\sigma_i^x) 
\end{equation}
with the ground state
\begin{equation}
  \ket{S} = \frac{1}{\sqrt{2^n}} \sum_{z=0}^{N-1} \ket z.
\end{equation}
However, this does not reflect the interaction topology of the final Hamiltonian \eqref{eq:final_ham}, i.e., the coefficient $n_i$ does not carry the full information about the clauses $c_i$. 

\subsubsection{XYZ-Algorithm}

A choice with two-qubit interactions is the Heisenberg ferromagnet \cite{schuetzhold2006b,schaller2010a}
\begin{equation}
  \label{eq:Hixyz}
  \Hi = \H^{xyz} = \Omega \sum_{i<j}^n \frac{n_{i,j}}{2} (\ID-\boldsymbol\sigma_i \cdot \boldsymbol\sigma_j).
\end{equation}
Both $\H^{xyz}$ and \Hf\ are invariant under rotations around the $\Sigma^z$-axis,
\begin{equation}
  \label{eq:Sigmaz}
  \Sigma^z = \sum_i^n \frac{1}{2} (\ID-\sigma_i^z).
\end{equation}
The eigenvalue $\Delta$ of $\Sigma^z$ is directly related to the number of $1$-bits in the solution 
(also denoted as Hamming weight $\Delta$).
It is therefore a constant of motion and conserved during dynamics. 
Only the subspace $\ket\psi:\Sigma^z \ket\psi = \Delta_w \ket \psi$ with the fixed Hamming weight $\Delta_w$ of the solution has to be considered here, 
as all subspaces with different Hamming weights
evolve independently.
The ground state of $\H^{xyz}$ in the appropriate subspace is given by a balanced superposition over all basis states $\ket u$ with  $\Sigma^z\ket u=\Delta_w\ket u$,
\begin{equation}\label{eq:initstate}
  \ket{\psi_0}^{xyz} = \bin{n}{\Delta_w}^{-\frac{1}{2}} \sum_u \ket u.
  % \begin{pmatrix} n \\ \Delta_w \end{pmatrix}    
\end{equation}
which can be prepared efficiently \cite{childs2002b} by adiabatic evolution $\H^x \overset T\rightarrow \H^P$. 
In that reference, the final Hamiltonian $\H^P$ 
used an energy penalty to separate the subspaces: It was given by $\H^P = \left( \Sigma^z-\Delta_w \right)^2$, 
which has highly degenerate energy levels, but due to symmetry arguments the correct angular momentum eigenstates were selected.
%GS
In our numerical considerations we circumvent this preparation step and directly prepare the initial state~(\ref{eq:initstate}), 
such that the adiabatic algorithm only
consists in a deformation of $\H^{xyz}$ to the final problem Hamiltonian \eqref{eq:final_ham}.

Realistically, the solution $\ket w$ and thus the Hamming weight $\Delta_w$ 
would not be known in advance, which would make repeated runs of the AQA in different 
subspaces necessary. 
But even in the worst case, when every possible value of $\Delta\in\{0\ldots n\}$ has to be tried, the computational overhead scales only linearly in $n$.

\subsubsection{XY-Algorithm}

Similar to the previous example is the x,y-ferromagnet
\begin{equation}
  \label{eq:Hixy}
  \Hi = \H^{xy} = 3m\Omega\ID -\Omega \sum_{i<j}^n \frac{n_{i,j}}{2} (\sigma_i^x\sigma_j^x+\sigma_i^y\sigma_j^y).
\end{equation}
It has already been shown numerically that it yields on average a better performance than $\H^{xyz}$ on a common 
instance of EC3 with a unique solution~\cite{schuetzhold2006b}.
Again, the Hamming weight is a constant of motion and we only consider the corresponding subspace. The ground state of (\ref{eq:Hixy}) is analytically unknown but can be initialized by adiabatic evolution 
$\H^{x} \overset T \rightarrow \H^{xy}$.
%GS
Using an ARPACK eigensolver~\cite{arpack1998,kandler2013a} accepting SQH as input variable, we found numerically 
for the examples considered that the minimum gap during the initial preparation is roughly independent 
of the system size.
We expect therefore only a mild algorithmic scaling of this preparation step with the system size $n$, which
would enable an efficient adiabatic preparation of the ground state of Eq.~\eqref{eq:Hixy}.

\subsection{Results}

The probability to find the system in the solution state~$\ket w$ after the runtime~$T$ is $P_{1}(T) = \abs{\braket{\psi(T)}{w}}^2$ and would ideally approach
one.
However, to avoid too long computation times but simultaneously ensure a high fidelity, we define a 
successful runtime $T_s$ by measuring the system's energy $E(T_s)/\Omega=\frac{1}{2}$. 
Here, we exploit the fact, that any excited state raises the energy by an amount greater or equal $\Omega$, 
cf. Eq.~\eqref{eq:final_ham}.
The first excited state of \Hf\ therefore has an energy $E_2\ge \Omega$.
The energy is thus lower bounded by (using $E_1(T)=0$)
\begin{eqnarray}
E(T) &=& \sum_{n=1}^N E_n(T) P_n(T) \ge E_2 (P_2+\ldots+P_N)\nonumber\\
&\ge& \Omega (1-P_{1})\,,
\end{eqnarray}
and defining this criterion as a measure for a successful runtime means that the ground state occupation is
at least one-half.

\begin{figure}[htb]
  \centering
  \includegraphics[width=\linewidth]{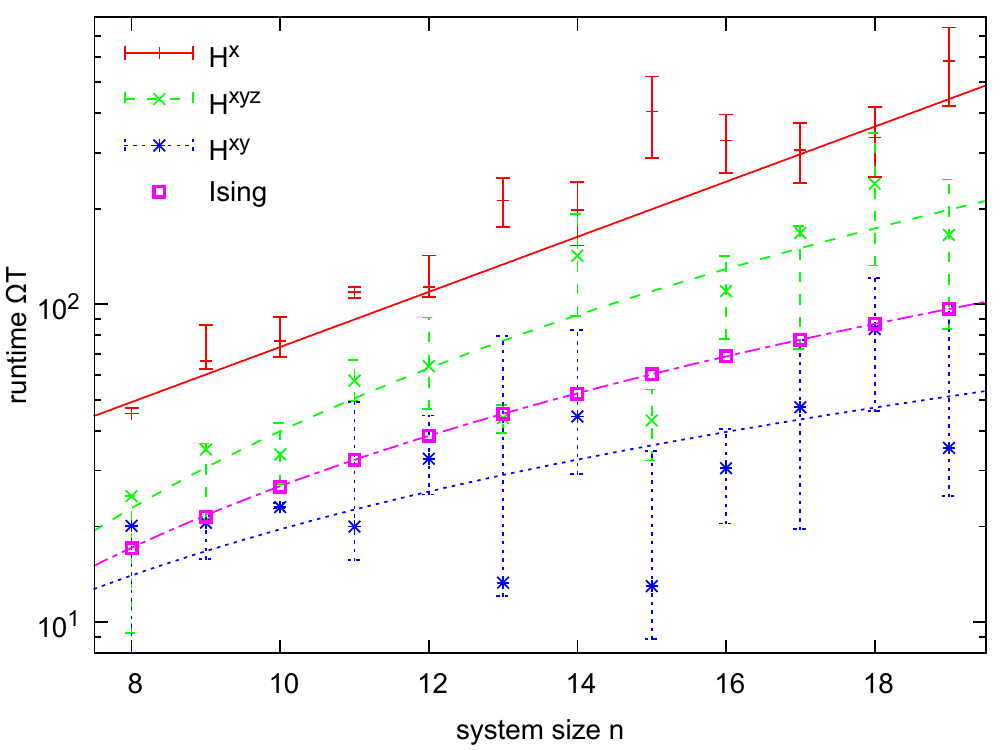}
  \caption{(Color Online) Scaling of the dimensionless runtime $\Omega T_s(n)$ (symbols, fit lines only serve to guide the eye) 
    for algorithms X (solid), XYZ (long-dashed), XY (dotted) compared with the quadratically scaling Ising model (dash-dotted).
    Symbols represent the median and error bars the first and third quartile out of 100 randomly chosen hard instances.
    Although the scaling behavior is inconclusive, it is evident that different initial states may
    drastically improve the performance of the AQA.  }
  \label{fig:Tscale}
\end{figure}

Figure \ref{fig:Tscale} shows computed median runtimes $T_s(n)$ for the three different algorithms compared with an adiabatic 
version of the Ising model, Eq.~(\ref{eq:ising}), traversed at constant interpolation speed $s(t)=t/T$.
The latter serves as a benchmark with a runtime scaling known to be quadratically \cite{dziarmaga2005a}. 
Our results show, that the XY-algorithm is the fastest followed by the XYZ-algorithm. Both stay close to the 
Ising curve which would indicate a polynomial scaling in the observed region. A clear statement however 
is hampered by the large deviations and fitting remains ambiguous. 
In contrast to previous studies \cite{farhi2001a}, where the runtime of the X-algorithm appeared 
polynomial on small system sizes, our results on harder instances suggest an exponential scaling of 
the original algorithm already for these moderate sizes $n$.

%%%%%%%%%%%%%%%%%%%%%%%%%%%%%%%%%%%%%%%%%%%%%%%%%%%%%
\section{Alternative Paths}\label{sec:alternative}
%%%%%%%%%%%%%%%%%%%%%%%%%%%%%%%%%%%%%%%%%%%%%%%%%%%%%

\begin{figure}[t]
  \centering
  \includegraphics[width=\linewidth]{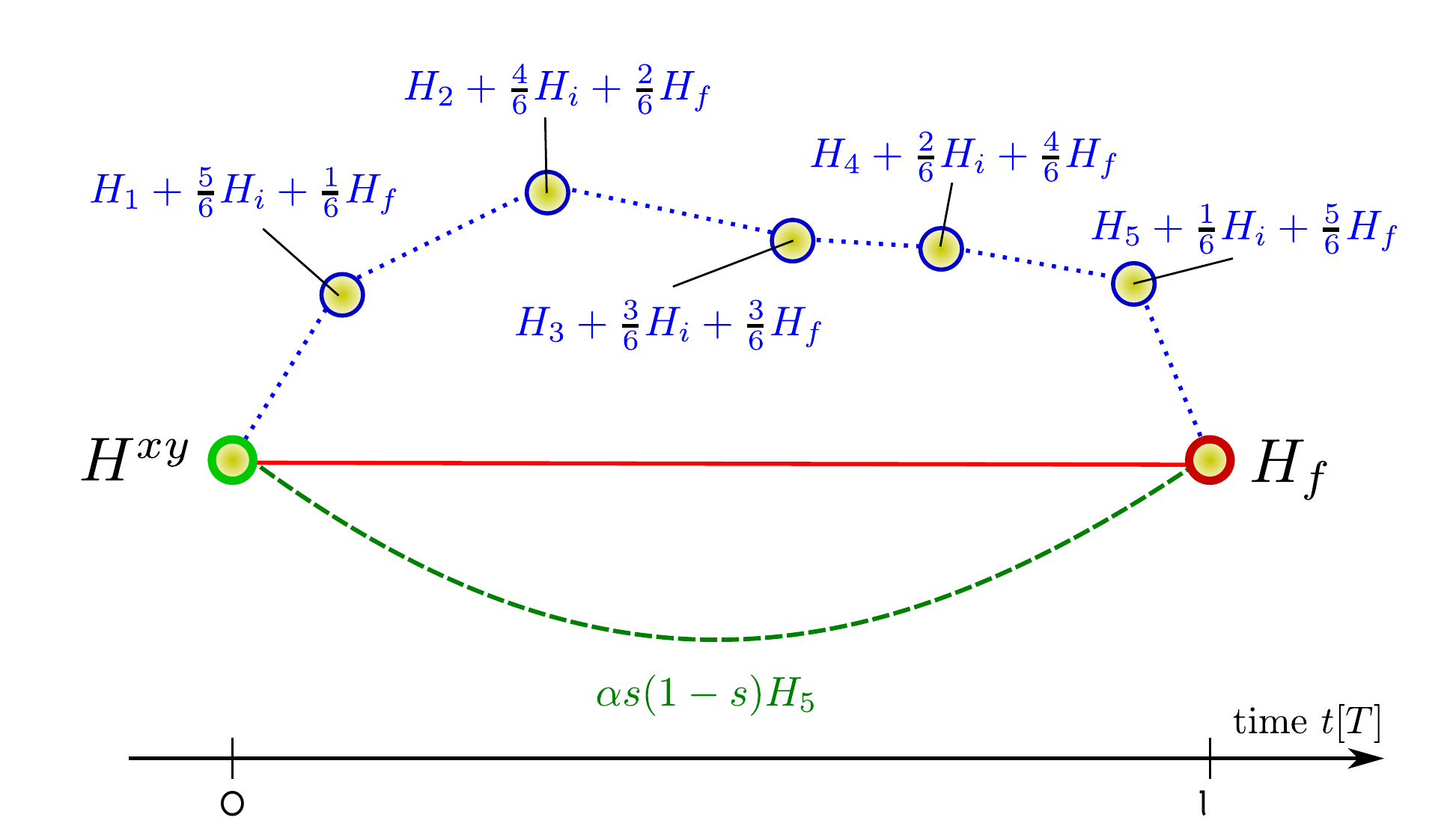}
  \caption{(Color Online) Sketch of alternative paths for adiabatic algorithms all starting from $\H^{xy}$. 
As an example, the number of clauses is set to $m=6$. The red, solid line is the straight line algorithm. The green,
  dashed curve denotes a path with an additional, nonlinear term $s(1-s)H_5$, which is defined in the text.
 A clause-by-clause algorithm is shown by the blue, dotted lines.}
  \label{fig:paths}
\end{figure}

A straight line interpolation between initial Hamiltonian \Hi\ and final Hamiltonian \Hf\ is a convex combination of only of two Hamiltonians. 
However, there are plenty of other paths connecting these two but involving a third or even more intermediate Hamiltonians. 
Our hope is, that some of such alternative paths may increase the energy gap above the ground
state leading to a speedup of the AQA. %review
For example, in the simple Ising model~(\ref{eq:ising}) it is known that even for constant-speed interpolations $s(t)=t/T$, 
the runtime can be improved from quadratic scaling (straight line) to linear scaling (nonlinear path)~\cite{schaller2008b}.
Since the XY-algorithm showed the best median
performance in the previous section, we set the initial Hamiltonian to $\H^{xy}$ and probe two 
alternative algorithms based on paths shown in Fig.~\ref{fig:paths}.

\subsection{Algorithms}

\subsubsection{Nonlinear Smooth Interpolation}

We add a third term to the straight line interpolating Hamiltonian $\H(s)$ in Eq.~\eqref{eq:aqc}, which is quadratic in $s$:
\begin{equation}
  \label{eq:nlpath}
  \H(s) = (1-s) \H^{xy} + s\Hf + \alpha s(1-s) \H_{m-1},
\end{equation}
%GS
where $\H_{m-1}=\Omega \sum_{i=1}^{m-1} h_i$ is the final Hamiltonian reduced by one (arbitrarily
chosen) clause  and $\alpha$ is a coupling strength.
To motivate this path, we note that for large $n$, the related reduced EC3 problem may be expected to have many 
solutions, since there exists a phase transition
from satisfiable to unsatisfiable EC3 problems at a clause-to-size ratio $m/n \approx 0.62$~\cite{kalapala0508037}.
Thus, reducing the number of clauses moves the problem into the satisfiable phase.
Intuitively, we expect that the additional term in Eq.~\eqref{eq:nlpath}, which becomes dominant during the evolution, 
will already at this state suppress states which are not a solution to $\H_{m-1}$ (and therefore neither of $\H_m$).
Thereby, the search space to find the solution of \Hf\ is reduced, and the algorithm could be expected to be faster 
compared to conventional straight line interpolation.

It should be noted, that for $m$ clauses there are $m$ different Hamiltonians $\H_{m-1}$. 
The best reduction of the search space is then obtained for reduced Hamiltonians $\H_{m-1}$ with the smallest number of solutions. 
However, this number will in realistic experiments not be known. 
In our numerical simulations, we have decided to 
remove only clauses when the connectivity of the graph is not destroyed.
In the example of Fig.~\ref{fig:ec3}, allowed clauses to be removed are $(1,4,13), (1,10,11)$ and $(9,11,13)$.

\subsubsection{Nonlinear Clause-By-Clause Interpolation}

We try again to reduce to search space by applying an additional term to the straight line interpolation. 
In contrast to the previous case however, the reduction is conducted not in a single step but by switching
 on the clauses one after another.  This can be written formally as
  \begin{align}
    \label{eq:cbc}
    \H(s) &= (1-s) \H^{xy} + s\Hf + \H_d(s), \\
    \label{eq:kpath}
    \H_d(s) &=  \sum_{k=1}^{m} 
    \begin{aligned}[t]
      &\big[ (1-s_k) \H_{k-1} + s_k\H_k \big] \\[1mm]
      &\times\Theta(s_k)\Theta(1-s_k),
    \end{aligned} \\[2mm]
    \text{where} \qquad s_k &= ms - k + 1.
  \end{align}
The primary interpolation $s:0\rightarrow1$ of Eq.~\eqref{eq:cbc} is thus split into $m$ steps,  
which consist of  secondary interpolations $s_k:0\rightarrow1$ from $\H_{k-1}$ to $\H_k$ in Eq.~\eqref{eq:kpath}. 
Here, $\H_k$ consists of $k$ clauses from \Hf. Note that this is equivalent to adding a single clause with each step denoted by $\H_k-\H_{k-1}$. 
As the initial and final Hamiltonians, $\H(0)=\H^{xy}$ and  $\H(1)=\Hf$ should not be changed by $\H_d$, we define $H_0 = \H_m = 0\cdot\ID$. 
The order, in which clauses are added, is ambiguous, which results in $m!$ possible paths.
In our numerical simulations, the path was defined by the order in which the clauses were stored.

\subsection{Results}

\begin{figure}[b]
  \centering
  \includegraphics[width=\linewidth]{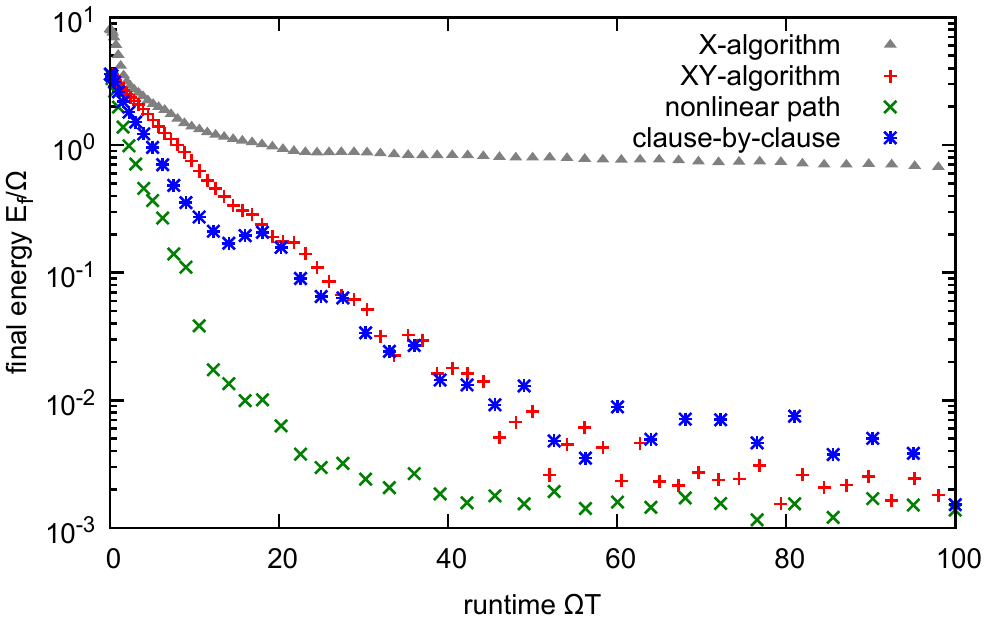} \vspace{2mm}
  \includegraphics[width=\linewidth]{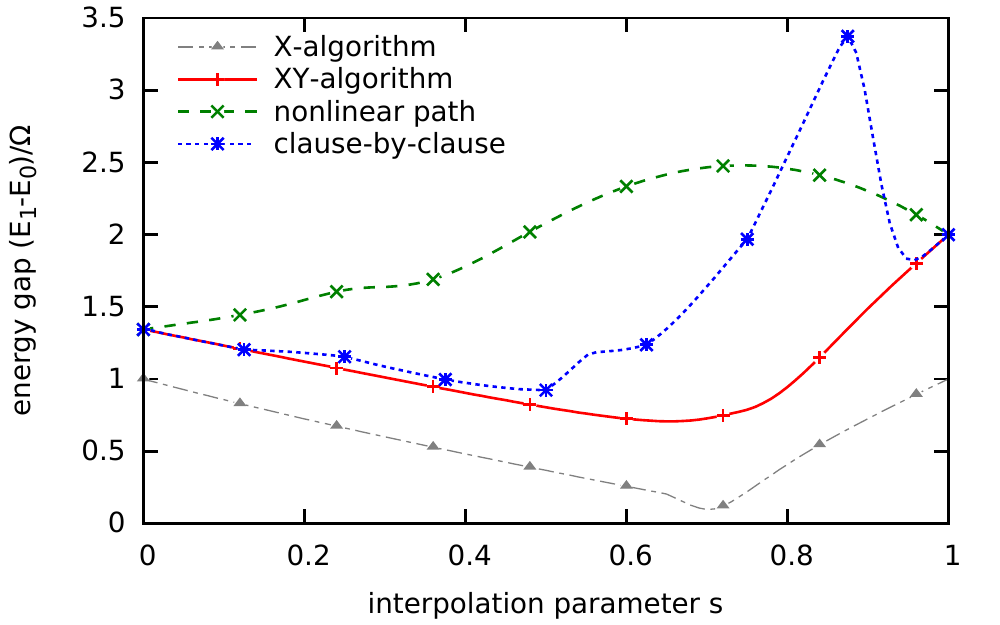}
  \caption{(Color Online) \textbf{Top:} Final energies versus dimensionless runtimes for different
    paths between $\H^{xy}$ and \Hf\ for a 13 qubit example.  Symbol shapes and colors correspond to
    the paths in Fig.~\ref{fig:paths}.  The X-algorithm is also shown as an example of a slow
    adiabatic behavior.  For the nonlinear path, the coupling is chosen as
    $\alpha=8$. \textbf{Bottom:} Energy gap between ground and first excited state of $\H(s)$ for
    the same algorithms. The symbols for the curves correspond to those in the upper panel. For the
    clause-by-clause algorithm, the symbols indicate the start and end of the secondary
    interpolations in \eqref{eq:kpath}. Comparing both panels, a correspondance between a large
    minimal gap and a fast decrease in the final energy is clearly seen.}
  \label{fig:41N13}
\end{figure}

The qubit system is prepared in the ground state of $\H^{xy}$, which can be done efficiently as
stated in section~\ref{sec:straight}.  We then compare for the presented algorithms the final energy
$E_f$ of the system after a runtime $T$.  For an adiabatic runtime, the energy should be close to 0.
Additionally, we numerically~\cite{arpack1998} compute the lower part of the spectrum and deduce the
energy gap above the ground state for each AQA. 
Figure~\ref{fig:41N13} exemplarily shows our results for an instance with 13 qubits. 
%GS
First, in the upper panel, it is visible that for short runtimes, the conventional algorithm
rapidly decreases its final energy, but it becomes increasingly hard to further reduce the energy
below the critical threshold one.
Both the clause-by-clause algorithm and the XY-algorithm decrease the final energy significantly
faster, but the nonlinear path shows an even better performance (we attribute the final plateaus to
imperfect numerical preparation of the initial ground state).

The lower panel in Fig.~\ref{fig:41N13} shows the gap between the two lowest eigenvalues of $\H(s)$
dependent of the interpolation parameter $s:0\rightarrow 1$. For the clause-by-clause algorithm, the
points indicate the particular steps where another clause is added. Again, the nonlinear paths shows
the best result as its minimum gap is largest. Comparing both panels, it can be clearly seen, that a
larger minimum gap leads to a faster decrease in the final energy. Remarkably, the minimum gap for
the nonlinear algorithm is located at $s=0$, i.e., it coincides with the gap of $\H^{xy}$.

\begin{figure}[tb]
  \centering
  \includegraphics[width=\linewidth]{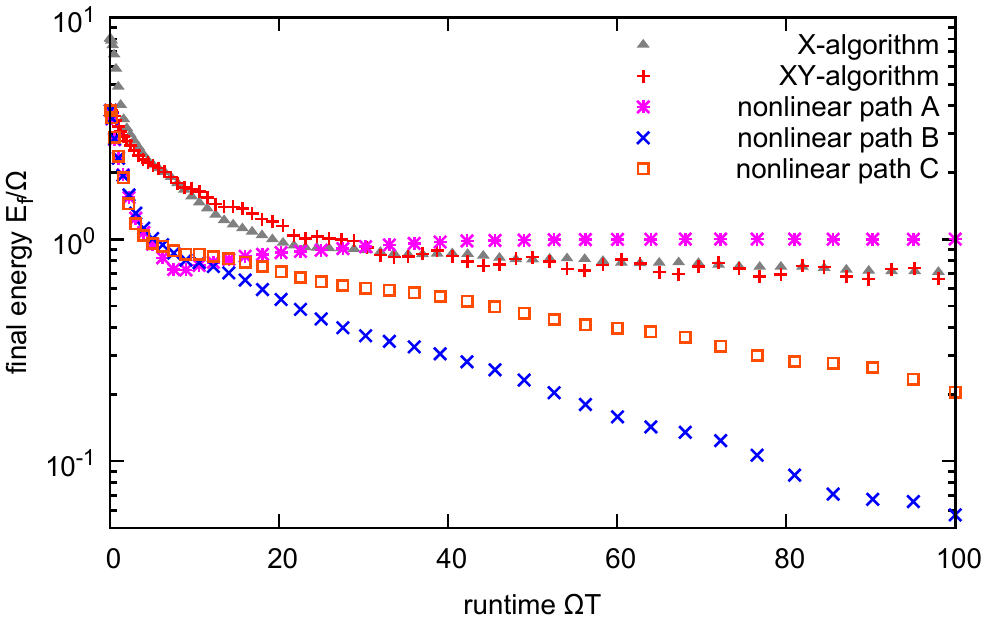} \vspace{2mm}
  \includegraphics[width=\linewidth]{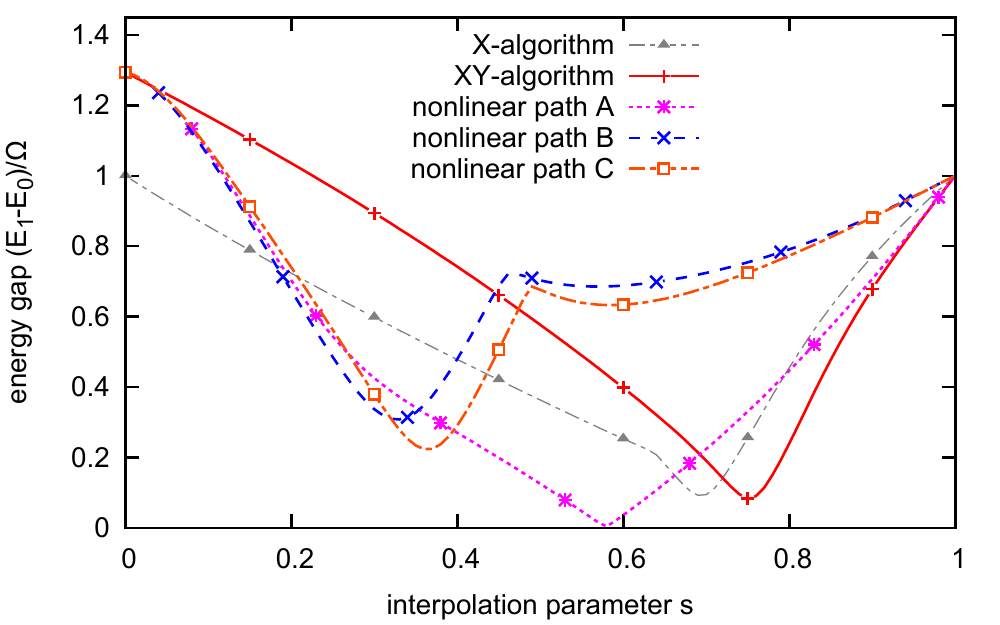}
  \caption{(Color Online) A very hard instance. \textbf{Top:} Final energy of the nonlinear path for different choices of
    $H_{m-1}$ with coupling $\alpha=8$ compared to the X- and XY-algorithm.  Clauses removed from
    \Hf\ for the nonlinear term are $A=(1,10,11),B=(1,4,13)$ and $C=(9,11,13)$.  Symbol shapes and
    colors correspond to the clauses in Fig.~\ref{fig:ec3}.  Whereas the short-time performance was
    much better for all nonlinear paths, it turned out that these differ strongly for large
    evolution times $T$, and choice $A$ performs worse than the straight line interpolations.
    \textbf{Bottom:} Energy gap for the same algorithms. Again, the value of the minimum gap clearly
    corresponds to the large-time performance in the above panel. It depends on the choice of
    $H_{m-1}$ whether this gap is increased or decreased compared to straight line algorithms.}
  \label{fig:1098N13}
\end{figure}

Although a broad statistics would exceed the scope of this paper, we observed a similar 
behavior for many instances and different system sizes.
However, there are very hard instances, where an arbitrarily chosen nonlinear algorithm failed.  The
graph depicted in Fig.~\ref{fig:ec3} is such an example, which is as hard to solve for the
XY-algorithm as it is for the X-algorithm.  In this case, our chosen smooth nonlinear path performed for a
large coupling constant $\alpha$ even worse.

The choice of the nonlinear term $H_{m-1}$ in \eqref{eq:nlpath} turns out to be crucial as shown in
Fig.~\ref{fig:1098N13}. In the upper panel, both straight line algorithms do not reduce the energy
significantly below 1. Also shown are the possible three nonlinear paths. Surprisingly, the energy
of path A is almost constant 1 for long runtimes. At least two out of three nonlinear paths show a
faster decrease in energy than the XY-algorithm, with path B having the best performance. This can
be confirmed by examining the energy gap in the lower panel of Fig.~\ref{fig:1098N13}. The minimum
gap of the straight line algorithms is almost identical, whereas the minimum gap of path A is 20
times smaller. Path B has the largest minimum gap as expected.

\section{Conclusion}

The employed universal integrator proved a flexible tool in simulating non-standard adiabatic
quantum algorithms.  The introduced SQH format offers an efficient storage scheme and a fast
matrix-vector multiplication.  Moreover, as the integrator is independent of the Hamiltonian's
structure, it gives the flexibility to simulate hundreds of EC3 instances without changing the
source code.  Adapting the integrator to alternative interpolating paths could be done very easily.

Our simulations agree with previous results pointing to an exponential scaling of the
X-algorithm~\cite{znidaric2005a} on hard instances.  The performance of the XYZ- and the
XY-algorithm is much better, indicating an Ising-like polynomial scaling for the samples and sizes
considered.  However, we note that variations are large, and the worst-case complexity does not even
expose any scaling behavior.
Even if this was not the case, finite-size simulations must remain inconclusive by construction.
For the alternative paths, we did not study the runtime scaling versus the problem size.
Instead, we considered the adiabatic behavior for exemplary instances, where a faster decrease in
the final energy corresponds to a larger minimum gap.
Here, the nonlinear path outperforms the linear algorithms even for very hard instances.
However, in general one will not know in advance, which of the possible choices for the nonlinear
path is the best.

For further studies, an analysis of its scaling behavior is of interest.  This requires extensive
simulations for statistics, which will be a subject of future research.

Our results are of course limited to the specific examples considered, but may give rise to the hope
that nonlinear paths may be an interesting road to explore in the field of adiabatic computation.

\section{Acknowledgments}

The authors have profited from discussions with C. Schr\"oder, U. Kandler, R. Okuyama, T. Brandes,
V. Mehrmann, R. Sch\"utzhold, and F. Renzoni.
Financial support by the DFG (BRA-1528/7-1, SCHA 1646/2-1) is also gratefully acknowledged.

%%%%%%%%%%%%%%%%%%%%%%%%%%%%%%%%%%%%%%%%%%%%%%%%%%%%%%%%%%%%%%%%%%%%%%%%
%%%%%%%%%%%%%%%%%%%%%%%%%%%%%%%%%%%%%%%%%%%%%%%%%%%%%%%%%%%%%%%%%%%%%%%%
%%%%%%%%%%%%%%%%%%%%%%%%%%%%%%%%%%%%%%%%%%%%%%%%%%%%%%%%%%%%%%%%%%%%%%%%
%%%%%%%%%%%%%%%%%%%%%%%%%%%%%%%%%%%%%%%%%%%%%%%%%%%%%%%%%%%%%%%%%%%%%%%%

%\bibliographystyle{unsrt}
\bibliography{paper_final}

\end{document}